\documentclass[aps,prx,twocolumn,showpacs,preprintnumbers,amsmath,amssymb,superscriptaddress]{revtex4-1}
\usepackage{amscd,amssymb,amsmath,latexsym}
\usepackage[mathcal,mathscr]{euscript}
\usepackage{amsfonts}
\usepackage{amsmath}
\usepackage{amssymb}
\usepackage{graphicx}
\usepackage{bm}
\usepackage{color}
\newcommand{\nc}{\newcommand}
\nc{\be}{\begin{equation}}
\nc{\ee}{\end{equation}}
\nc{\bea}{\begin{eqnarray}}
\nc{\eea}{\end{eqnarray}}
\nc{\bean}{\begin{eqnarray*}}
\nc{\eean}{\end{eqnarray*}}
\nc{\mb}{\mbox}
\nc{\rnc}{\renewcommand}
\nc{\vk}{\mb{\bmk}}
\nc{\vp}{\mb{\bmp}}
\nc{\vn}{\mb{\bmn}}
\nc{\vq}{\mb{\bmq}}
\nc{\rr}{\mb{\bmr}}
\nc{\vz}{\hat {\mb{\bmz}}}
\nc{\vj}{\mb{\boldmath$j$}}
\nc{\vg}{\mb{\boldmath$g$}}
\nc{\x}{\mb{\boldmath$x$}}
\nc{\A}{\mb{\boldmath$A$}}
\nc{\va}{\mb{\boldmath$a$}}
\nc{\vs}{\mb{\boldmath$\sigma$}}
\nc{\vpi}{\mb{\boldmath$\pi$}}
\nc{\nab}{\nabla}
\nc{\X}{\sf x}
\usepackage{txfonts}
\usepackage{pxfonts}
\usepackage{graphicx,bm,units,yfonts}
\usepackage{hyperref}

\newcommand{\BM}{{\mathbb B}}
\newcommand{\CM}{{\mathbb C}}

\newcommand{\IM}{{\mathbb I}}

\newcommand{\RM}{{\mathbb R}}

\newcommand{\ZM}{{\mathbb Z}}

\newcommand{\Kk}{{\mathcal K}}

\newcommand{\Ll}{{\mathcal L}}

\begin{document}

\title{Topological Classification Table Implemented with Classical Passive Meta-Materials}


\author{Yafis Barlas and Emil Prodan}

\affiliation{Department of Physics, Yeshiva University, New York, NY, USA}

\begin{abstract}
Topological condensed matter systems from class A and class AII of the classification table have received classical electromagnetic and mechanical analogs and protected wave-guiding with such systems has been demonstrated experimentally. Here we introduce a map which generates classical analogs for all entries of the classification table, using only passive elements. Physical mechanical models are provided for all strong topological phases in dimension 2, as well as for three classes in dimension 3. This includes topological super-conducting phases, which have never been attempted with classical systems.
\end{abstract}

\maketitle

\section{Introduction}

Topological methods \cite{ThoulessPRL1982,HaldanePRL1988} have revolutionized the way we approach the design of our materials and the functionalization of our devices. These days, condensed matter, photonic and mechanical/acoustic crystals with demonstrated intrinsic topological properties are common. One of their common characteristics is the spontaneous emergence of wave-guiding modes whenever a boundary is cut into a material and, for strong topological materials, these modes are robust against disorder. The strong topological condensed matter systems were classified at the end of the previous decade \cite{SRFL2008,QiPRB2008,Kit2009,RSFL2010} and it emerged that there are only three fundamental symmetries that can stabilize these phases, namely, the time-reversal (TR), particle-hole (PH) and chiral (CH) symmetries. Their combinations led to the 10 distinct topological classes reproduced in Table~\ref{Tab-ClassTable}. 

About the same time, the question of whether such exotic physical properties can be reproduced with classical systems started circulating within the physics community. For class A in dimension $d=2$, the affirmative answer was provided in \cite{HaldaneRaghu2008,PP2009} for photonic and mechanical systems, respectively, and experimental confirmations soon followed \cite{WangNature2009,NashPNAS2015}. Rapid progress happened afterwards. Electromagnetic \cite{HafeziNatPhot2013,WuPRL2015} as well as mechanical \cite{SusstrunkScience2015} systems exhibiting the physics of AII class in $d=2$ have been predicted and realized in laboratories. Classical mechanisms inspired from AIII class has been demonstrated in \cite{KaneNatPhys2013,PauloseNatPhys2015} and a linear mechanical chain from BDI class with locked-in particle-hole symmetry and Majorana edge excitations was observed in \cite{ProdanNatComm2017}. An electromagnetic emulation of a topological insulator from AII class in $d=3$ was proposed in \cite{KhanikaevNatPhot2017} and mechanical systems that emulate the quantum spin-Hall effect with passive materials have also been proposed \cite{MousaviNatComm2015} and implemented \cite{RuzzeneArxiv2017,ChaunsaliPRB2018,ChernArxiv2018}.

Among all, the S{\"u}sstrunk-Huber technique \cite{SusstrunkScience2015} and its generalization \cite{PalJAP2016} stand out because they involve only passive components and the TR-like symmetry that stabilizes the edge modes is global and exact. In contradistinction, in all the other approaches based on passive components, such as \cite{WuPRL2015} or \cite{MousaviNatComm2015}, this symmetry only holds at one point or between two points of the Brillouin torus, hence TRS protection occurs only in a small frequency interval where a certain effective model applies. 

The mapping from the class A to class AII devised in \cite{SusstrunkScience2015,PalJAP2016} carries an inherent additional $U(1)$ symmetry, hence the output is always a spin-Chern system with topological phases classified by $\ZM$ instead of $\ZM_2$. This is unfortunate, for there is unique physics associated to latter that cannot be probed with spin-Chern insulators. For example, the generic systems from AII-class in $d=2$ can support metallic phases in the presence of disorder, but such phases vanish when an additional $U(1)$-symmetry is present. For this reason, the critical quantum regimes at the topological transitions occuring in class AII are fundamentally different from the ones supported by spin-Chern insulators. Also, the protection mechanisms of the edge modes against Anderson localization are fundamentally different for these two classes of systems.

The most acute shortcoming of that map is its inability to produce topological  classical systems from AII class in $d=3$, where the analog of spin-Chern insulator does not exist. For the rest of the topological classes, the situation is even worse because no such map has ever been attempted. For example, classical analogs of topological systems from any of the BdG classes in dimension higher than one do not exist.  Let us point out again that the distinct entries in the classification table implement the ten universal disordered classes, each of them displaying unique characteristics \cite{ZirnbauerJMP1996, AltlandPRB1997}, and many have never been observed experimentally. Furthermore, each topological class displays unique bulk and boundary physical responses, which may enable important physical applications \cite{HasanRMP2010}. As such, despite the recent progress with classical systems, there are many missed opportunities, which we hope can be addressed in the near future with the tools provided by our present work.

Indeed, we demonstrate here how the map \cite{SusstrunkScience2015,PalJAP2016} can be reformulated and generalized to ultimately cover the whole Table~\ref{Tab-ClassTable}. The final outcome is an algorithmic procedure to translate any strong topological condensed matter system from the classification table into an absolutely equivalent topological classical meta-material with passive components. More precisely, we will show that, if the waves in the proposed classical meta-material are excited in a prescribed way, then their propagation is determined by the Schrodinger equation of the corresponding strong topological condensed matter system. To exemplify, we generate explicit mechanical analogs of all topological classes in $d=2$, as well as three additional  classes in $d=3$, that have never been attempted before, classically. 

Let us stress that passive meta-materials do not require any external intervention to function at the desired parameters and, for this reason, their design can be easily scaled up or down if needed. As observed in the original work \cite{PP2009}, Lorentz-type forces, {\it i.e.} those involving the velocity, add complex entries in the dynamical matrix of the classical wave propagation and, for this reason, any tight-binding quantum Hamiltonian can, in principle, be reproduced by such classical dynamical matrix. In the passive meta-materials, the Lorentz forces are absent and, for this reason, emulating the strong spin-orbit coupling responsible for the topology in the condensed matter systems, is difficult. This is the challenge we overcome in our work. To give an intuitive picture of our solution, let us consider a topological phase from class A, which requires the breaking of the time-reversal symmetry. In passive meta-materials, the time-reversal symmetry is always present but it can be broken by the initial conditions. If we do so in a prescribed way and in our proposed meta-materials, then the solution of the wave equation propagates in a dynamically invariant sub-space where the time-reversal symmetry is broken. It is in this invariant subspace where the topological phase will be observed with the classical meta-material. This will be more or less the picture for all topological phases from the table. The extraordinary claim of our work is that, on this dynamically invariant sub-space, the time-reversal symmetry can be reinstated using passive couplings and same applies to the particle-hole symmetry.

\begin{table}\label{Table1}
{\scriptsize
\begin{center}
\begin{tabular}{|c|c|c|c||c||c|c|c|c|c|c|c|c|}
\hline
$j$ & TRS & PHS & CHS & CAZ & $d=0,8$ & $d=1$ & $d=2$ & $d=3$ & $d=4$ & $d=5$ & $d=6$ & $d=7$
\\\hline\hline
$0$ & $0$ &$0$&$0$& A  & $\ZM$ &  & \textcolor{red}{$\ZM$} &  & $\ZM$ &  & $\ZM$ &  
\\
$1$& $0$&$0$&$ 1$ & AIII & & $\ZM$ &  & \textcolor{red}{$\ZM$}  &  & $\ZM$ &  & $\ZM$
\\
\hline\hline
$0$ & $+1$&$0$&$0$ & AI &  $\ZM$ & &  & & $2 \, \ZM$ & & $\ZM_2$ & ${\ZM_2}$
\\
$1$ & $+1$&$+1$&$1$  & BDI & $\ZM_2$ &$\ZM$  & &  &  & $2 \, \ZM$ & & $\ZM_2$
\\
$2$ & $0$ &$+1$&$0$ & D & $\ZM_2$ & ${\ZM_2}$ & \textcolor{red}{$\ZM$} &  & & & $2\,\ZM$ &
\\
$3$ & $-1$&$+1$&$1$  & DIII &  & $\ZM_2$  & \textcolor{red}{$\ZM_2$} &  \textcolor{red}{$\ZM$} &  & & & $2\,\ZM$
\\
$4$ & $-1$&$0$&$0$ & AII & $2 \, \ZM$  & &  \textcolor{red}{$\ZM_2$} & \textcolor{red}{${\ZM_2}$} & $\ZM$ & & &
\\
$5$ & $-1$&$-1$&$1$  & CII & & $2 \, \ZM$ &  & $\ZM_2$  & $\ZM_2$ & $\ZM$ & &
\\
$6$ & $0$ &$-1$&$0$ & C&  &  & \textcolor{red}{$2\,\ZM$} &  & $\ZM_2$ & ${\ZM_2}$ & $\ZM$ &
\\
$7$ & $+1$&$-1$&$1$  &  CI &  & &   & $2 \, \ZM$ &  & $\ZM_2$ & $\ZM_2$ & $\ZM$
\\
[0.1cm]
\hline
\end{tabular}
\end{center}
}
\caption{\small Classification table of strong topological insulator and superconductors \cite{SRFL2008,QiPRB2008,Kit2009,RSFL2010}, listing all strong topological condensed matter phases according to their fundamental symmetries and space dimension. The phases highlighted in red will be explicitly mapped into classical systems.}
\label{Tab-ClassTable}
\end{table}  

Let us conclude that, in contradistinction, the active meta-materials require continuous pumping of energy to function and Lorentz forces can be introduced using different strategies, such as with the use of gyroscopes \cite{NashPNAS2015}. For these reasons, the emulation of strong spin-orbit couplings is easier with active meta-materials but scaling down the designs, {\it e.g.} to micron and nano scales, is difficult if not impossible.

\section{Coupled Mechanical Resonators}
\label{Sec:MechRez}

\begin{figure}
\includegraphics[width=\linewidth]{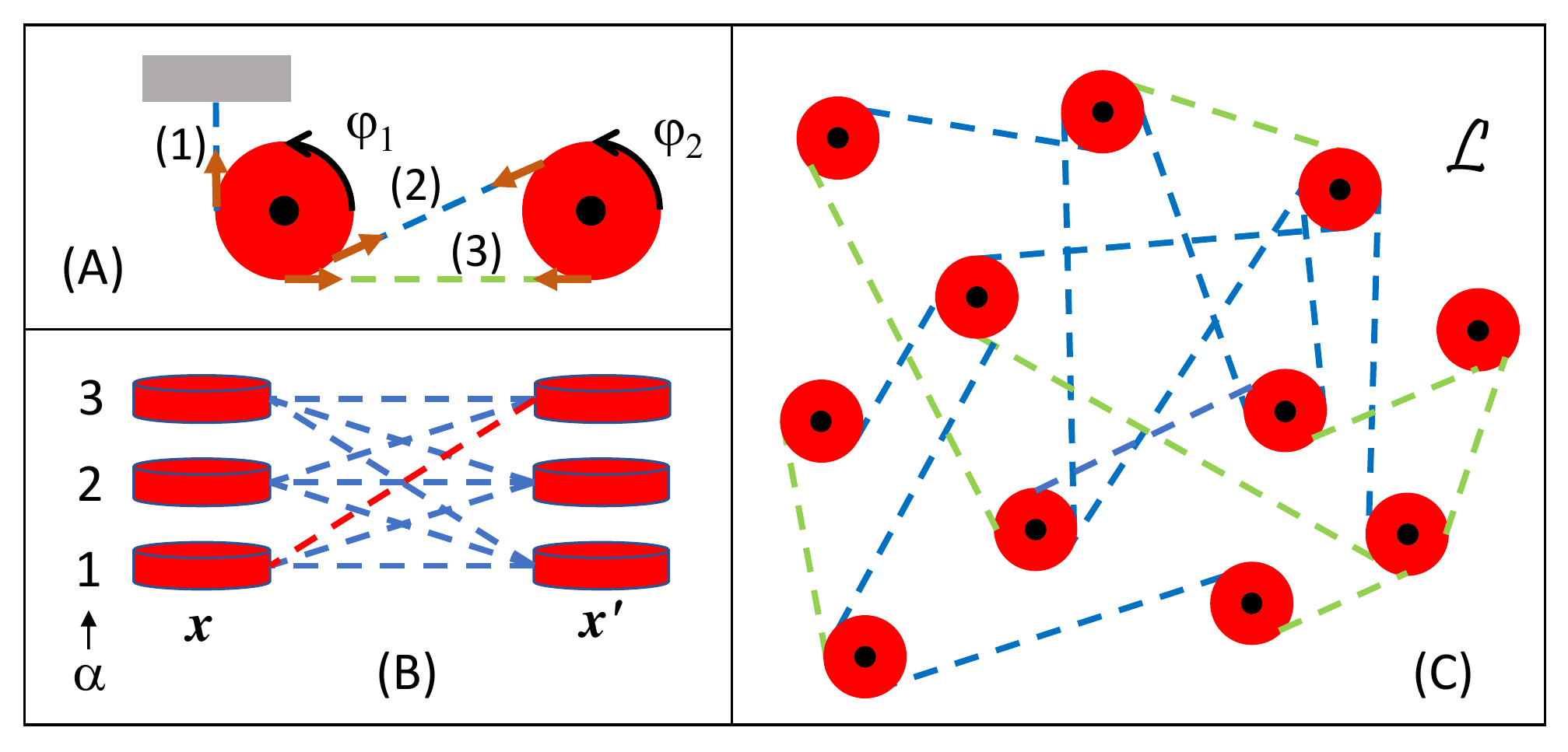}
\caption{\small (A) Spinners with fixed center of mass and one rotational degree of freedom couple via attractive forces (see arrows) to: (1) a fixed massive object or (2-3) a neighboring spinner, leading to two-body potentials $V \approx \tfrac{1}{2} a_1 \varphi_1^2 + \tfrac{1}{2} a_2 \varphi_2^2 + b \varphi_1 \varphi_2$, with $b$ positive (2) or negative (3). The $b$ coefficients are represented by connecting dotted lines while all $a$'s are incorporated in lines of type (1). (B) Spinners can be stacked and resonators with multiple degrees of freedom can be engineered and easily connected. These connections are directly related to the hopping matrices $\hat d_{\bm x,\bm x'}$. For example, the strength of the red-dashed connection provides the hopping coefficients $d_{\bm x,\bm x'}(1,3)$, as discussed in the text. (C) A generic lattice of coupled resonators. Note the color code, which will be applied throughout.}
\label{Fig:MechanicalSystem}
\end{figure}

We consider a collection of coupled passive resonators placed on a lattice $\Ll$, such as the one described in Fig.~\ref{Fig:MechanicalSystem}, with the following essential features: 1) existence of $N$ degrees of freedom $q^\alpha_{\bm x}$ localized around the vertices $\bm x$ of $\Ll$ and 2) the ability to couple $q^\alpha_{\bm x}$'s, one pair at a time (see \cite{ApigoArxiv2018,QianArxiv2018} for a laboratory realization). In this setting, the configuration space consists of column matrices $\bm Q=\{q^\alpha_{\bm x} \}_{\bm x \in \Ll}^{\alpha=\overline{1,N}}$ and, in the regime of small oscillations relative to a stable equilibrium point, the dynamics is determined by a quadratic Lagrangian $L = \tfrac{1}{2}\dot{\bm Q}^T \hat{\bm T} \dot{\bm Q} - \tfrac{1}{2}{\bm Q}^T \hat{\bm W} {\bm Q}$, leading  to the equations of motion $\ddot {\bm Q} = - D \bm Q$, with the dynamical matrix given by $D=\hat{\bm T}^{-\frac{1}{2}} \hat{\bm W} \hat{\bm T}^{-\frac{1}{2}}$. Since all the topological examples can be constructed with identical resonators, we will assume that $\hat{\bm T}$ is diagonal, hence the dynamical matrix can be read-off directly from our diagrammatic models, as explained below.

Due to our assumptions, the equations of motion have a particular structure. Indeed, $\bm Q$ can be viewed as a function $\bm Q:\Ll \rightarrow \CM^N$, $\bm Q(\bm x) = (q^1_{\bm x},\ldots,q^N_{\bm x} )^T$, and this space of functions can be organized as a complex Hilbert space, usually denoted by $\ell^2(\Ll,\CM^N) \simeq \CM^N \otimes \ell^2(\Ll,\CM)$, via the scalar product $\langle \bm Q,\bm Q' \rangle = \sum_{\bm x \in \Ll} \bm Q_{\bm x}^\dagger \cdot \bm Q'_{\bm x}$. With the ansatz $\bm Q(t) = {\rm Re}\big [ e^{\imath \omega t} \bm Q \big]$, all the resonant modes of the system can be found by solving the eigen-system $\omega^2 \bm Q = D \bm Q$ on $\ell^2(\Ll,\CM^N)$. On this particular Hilbert space, the most general form of a dynamical matrix takes the form:
\begin{equation}\label{Eq:GenDynMatrix}
D = \sum_{\bm x,\bm x' \in \Ll} \hat d_{\bm x,\bm x'} \otimes |\bm x \rangle \langle \bm x' |, \quad \hat d_{\bm x,\bm x'} = \hat d_{\bm x',\bm x}^T  \in M_{N \times N}(\RM), 
\end{equation}
where $|\bm x\rangle$ denotes the standard orthonormal basis of $\ell^2(\Ll,\CM)$, $\psi_{\bm x}(\bm x') = \delta_{\bm x,\bm x'}$. Throughout, we will use $\ell^2(\Ll)$ instead of $\ell^2(\Ll,\CM)$ and we will also denote by $\xi_1, \ldots, \xi_N$ the standard basis of $\CM^N$. Since we avoid active materials, the entries of $\hat d$'s are all real and they can be easily read-off from diagrams like Fig.~\ref{Fig:MechanicalSystem}(B). For example, $d_{\bm x,\bm x'}(1,3) = d_{\bm x',\bm x}(3,1)$ is equal to the numerical value associated to the red connection in Fig.~\ref{Fig:MechanicalSystem}(B). In the concrete examples, these values will be placed explicitly on the connecting lines or specified in the captions. Reciprocally, once given the $\hat{d}$ matrices, one can quickly sketch a physical representation of the mechanical system driven by \eqref{Eq:GenDynMatrix}, by reversing the process we just described. Because of these reasons, while somewhat mathematical, we find the notation in \eqref{Eq:GenDynMatrix} extremely useful. Let us also mention that if the system is translational invariant, then:
\begin{equation}\label{Eq:TranslationalH}
D = \sum_{\bm q}\sum_{\bm x \in \ZM^d} \hat d_{\bm q} \otimes |\bm x  + \bm q \rangle \langle \bm x| = \sum_{\bm q} \hat d_{\bm q} \otimes S_{\bm q},
\end{equation} 
with $S_{\bm q}|\bm x \rangle = |\bm x  + \bm q \rangle$, from where the momentum representation $\hat D(\bm k)= \sum_{\bm q} e^{\imath \bm q \cdot \bm k} \, \hat d_{\bm q}$, $\bm k \in [-\pi,\pi]^d$, follows.

\section{The Algorithm}
\label{Sec:Algorithm}

We start from a generic hermitian Hamiltonian $H$ on the complex Hilbert space $\CM^N \otimes \ell^2(\Ll)$, which necessarily takes the form \eqref{Eq:GenDynMatrix} but the hopping matrices are not real in general:
\begin{equation}\label{Eq:OrigHam}
H = \sum_{\bm x,\bm x' \in \Ll} \hat h_{\bm x,\bm x'} \otimes |\bm x \rangle \langle \bm x' |, \quad \hat h_{\bm x,\bm x'} = \hat h_{\bm x,\bm x'}^\dagger \in M_{N \times N}(\CM).
\end{equation}
Any model Hamiltonian from the classification table of topological condensed matter systems, periodic or disordered, can be written as in \eqref{Eq:OrigHam}. The symmetries induce certain additional constraints on $\hat h$-matrices.

{\em The map defined.} Let $\Kk$ be the complex conjugation on the space $\CM^N \otimes \ell^2(\Ll)$. Since the standard basis of $\ell^2(\Ll)$ is real, $\Kk$ acts on the Hamiltonians as $\hat h_{\bm x,\bm x'} \rightarrow \hat h_{\bm x,\bm x'}^\ast$, 
where $\ast$ stands for the ordinary complex conjugation. Now, let $\BM\big [\CM^N \otimes \ell^2(\Ll)\big ]$ denote the algebra of continuous linear operators over $\CM^N \otimes \ell^2(\Ll)$, of which $H$ is part of. We are going to define an algebra morphism:
\begin{equation} 
\rho: \BM\big [\CM^N \otimes \ell^2(\Ll)\big ] \rightarrow \BM\big [\CM^{2N} \otimes \ell^2(\Ll)\big ],
\end{equation} 
such that $\Kk^{-1} \rho(A) \Kk = \rho(A)$ for arbitrary and possibly non-hermitean $A=\sum_{\bm x,\bm x' \in \Ll} \hat a_{\bm x,\bm x'} \otimes |\bm x\rangle \langle \bm x' |$. Explicitly:
\begin{equation}
\rho(A) = \sum_{\bm x,\bm x' \in \Ll} \begin{pmatrix} {\rm Re}[\hat a_{\bm x,\bm x'}] & {\rm Im}[\hat a_{\bm x,\bm x'}] \\ -{\rm Im}[\hat a_{\bm x,\bm x'}] & {\rm Re}[\hat a_{\bm x,\bm x'}] \end{pmatrix} \otimes |\bm x \rangle \langle \bm x' |.
\end{equation}
It is straightforward to verify that $\rho$ respects the algebraic operations, {\it e.g.} $\rho(AB)=\rho(A)\rho(B)$, sends the identity operator into the identity operator, $\rho(\IM) = \IM$, respects the operation of taking the adjoint, $\rho(A^\dagger)=\rho(A)^\dagger$, and is into, $\rho(A) = \rho(B)$ if and only if $A=B$. Among many things, the properties we just listed ensure that the spectral properties are preserved \cite{DavidsonBook}:
\begin{equation}\label{Eq:SpecTransf}
{\rm Spec}\big ( \rho(H) \big ) = {\rm Spec}(H).
\end{equation} 
Let us point out that, when applied on the Hofstadter model with with $1/3$ flux quanta per cell, the map returns the mechanical system analyzed in \cite{SusstrunkScience2015}.

\noindent{\em Transfer of symmetries.} Consider the transformation: 
\begin{equation}
U: \CM^{2N}\otimes \ell^2(\Ll) \rightarrow \CM^{2N}\otimes \ell^2(\Ll), \quad U = {\scriptsize \begin{pmatrix} 0 & \IM_N \\ -\IM_N & 0 \end{pmatrix}} \otimes \IM, 
\end{equation}
which satisfies $ U^{-1} =U^\dagger = U^{T}= -U$ and hence $U^2=-1$. This implies that ${\rm Spec}(U) = \{\pm \imath\}$ and the spectral projectors on the eigen-subspaces are given by:
\begin{equation}
\Pi_\pm = \tfrac{1}{2}(1 \mp \imath U) = \tfrac{1}{2} {\scriptsize \begin{pmatrix} \IM_N & \mp \imath \, \IM_N \\ \pm \imath \, \IM_N & \IM_N \end{pmatrix}} \otimes \IM.
\end{equation}
It is immediate to verify that $[\rho(H), U] = 0$, hence $\rho(H)$ has automatically a built-in $U(1)$ symmetry. Furthermore, $\rho(H) = \Pi_- \rho(H) \Pi_- \oplus \Pi_+\rho(H) \Pi_+$, and $\Pi_\pm \rho(H) \Pi_\pm$ are unitarily equivalent to $H$ and $H^\ast$, respectively. Note that $\Kk \Pi_\pm \Kk^{-1} = \Pi_\mp$ and if:
\begin{equation}
J: \CM^{2N}\otimes \ell^2(\Ll) \rightarrow \CM^{2N}\otimes \ell^2(\Ll), \quad J = {\scriptsize \begin{pmatrix} 0 & \IM_N \\ \IM_N & 0 \end{pmatrix}} \otimes \IM, 
\end{equation}
then $J \rho(H) J^{-1} = \rho(H^\ast)$ and $J \Pi_\pm J^{-1} = \Pi_\mp$.
 
 Since $\rho(H)$ is real, it also has a built-in fermionic TR-symmetry $\Sigma \rho(H) \Sigma^{-1} = \rho(H)$, where $\Sigma = U \Kk = \Kk U$, $\Sigma^2 = - 1$. As such, Kramer pairing is automatically present for the classical system and the spectrum is necessarily doubly degenerate. It is important to note that this time-reversal symmetry and the $U(1)$ symmetry derived from $U$ are always simultaneously present. As such, the map $\rho$, just by itself, cannot generate systems from $AII$ class classified by a $\ZM_2$ invariant. It then becomes clear that \cite{SusstrunkScience2015} produced a spin-Chern insulator rather than a true quantum spin-Hall insulator. The conclusion is that $\Sigma$ is not the anti-unitary operation that will ultimately stabilize the true topological phases from AII class. Indeed, as we shall demonstrate next, we can correctly transfer all the symmetries, hence generate all the topological strong phases, if we restrict to the $\Pi_\pm$ sectors.
 
 Henceforth, assume that $\Theta H \Theta^{-1} =\epsilon H$, $\epsilon=\pm 1$, with $\Theta$ anti-unitary and $\Theta^2 = \gamma$, $\gamma= \pm 1$, and recall that, always, such $\Theta$ can be written as $\Kk W = W \Kk$ with $W$ unitary. The map $\rho$ applies only on linear operators, hence we do not know, as of yet, how to transfer anti-unitary symmetries. Nevertheless, based on the remarkable identity:
  \begin{equation}
	\label{Eq:mappedsymmetry}
\epsilon \, \Pi_\pm \rho(H) \Pi_\pm = \big (\Kk \rho(W) J\big ) \Pi_\pm\rho(H) \Pi_\pm \big ( \Kk \rho(W) J \big )^{-1},
 \end{equation}
we find that the proper mapping of $\Theta$ is:
\begin{equation}\label{Eq:Signature}
\Theta \rightarrow \tilde \Theta = \Kk \, \rho(W) J, \quad \tilde \Theta^2 =\gamma.
\end{equation}
The proof of the statement is supplied in the supplemental material. At this point, we reached the important conclusion: {\em If the original Hamiltonian enjoys any of the symmetries in the classification table, then the mapped Hamiltonian does too when projected on either of the $\Pi_\pm$ sectors.} 

The dynamical matrix $D$ in Eq.~\ref{Eq:GenDynMatrix} must satisfy one more constraint, namely, its spectrum needs to be positive. As such, we will take $D = \rho(H) + E_0$, $E_0 = \inf\{{\rm Spec}(\rho(H))\}$. This provides the coupling matrices $\hat d_{\bm x,\bm x'}$ which, at their turn, provide the explicit mechanical models, as explained in section~\ref{Sec:MechRez}.

\begin{figure}
\begin{center}
\includegraphics[clip,width=\linewidth]{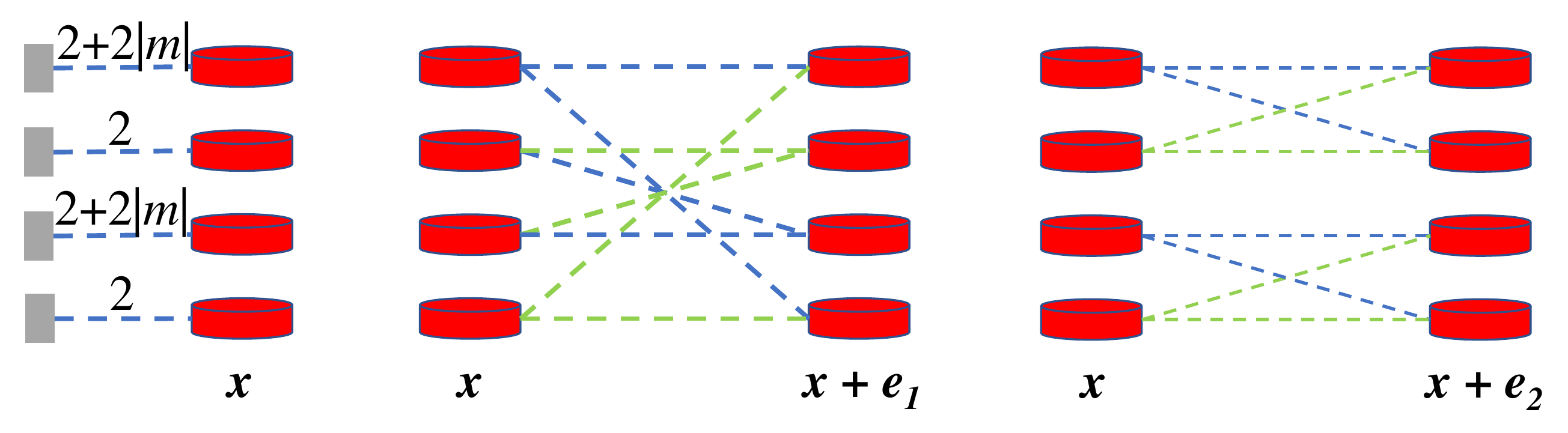}
\caption{\small Schematic representation of the elementary couplings in the realization of the dynamical matrix \eqref{Eq:MappedExample1} corresponding to the generating model of the D-A class. In the figure, $\bm e_j$'s are the primitive vectors of the Bravais lattice in 2D. All couplings have the strength $b=\pm 0.5$, with blue(green) denoting the positive(negative) coupling. The on-site coupling is indicated first panel. The edge spectrum for this model in the topological phase is shown in Fig.~\ref{Fig:2DEdgeSpectra} (a) \& (b).} 
\label{Fig:MappedHam1}
\end{center}
\end{figure}
 
\section{Bulk-Boundary Correspondence}

We divide the lattice in lef/right regions $\Ll = \Ll_L \cup \Ll_R$ and we restrict $H$ on the Hilbert space $\CM^N \otimes \ell^2(\Ll_R)$ via:
\begin{equation}
\widehat H = \sum_{\bm x,\bm x' \in \Ll_R} \hat h_{\bm x, \bm x'} \otimes |\bm x \rangle \langle \bm x' |.
\end{equation}
By doing so, we cleanly sever all the connections between the resonators in sub-lattice $\Ll_R$ and the ones in sub-lattice $\Ll_L$, but our arguments apply for more general boundary conditions, as we shall see later. This operation defines a linear map:
\begin{equation}\label{Eq:CutHam}
{\rm Cut} : \BM\big [\CM^N \otimes \ell^2(\Ll) \big ] \rightarrow \BM\big [\CM^N \otimes \ell^2(\Ll_R) \big ],
\end{equation} 
which in general does not respect the multiplication. A similar map can be defined on the Hilbert spaces with doubled internal degrees of freedom. Now, note that $\rho$ acts ultra-locally, in the sense that the internal degrees of freedom at a point $\bm x$ are not mixed with the ones at another point $\bm x'$. Due to this ultra-locality, $\rho \circ {\rm Cut} = {\rm Cut} \circ \rho$ and, since the original lattice was arbitrary, \eqref{Eq:SpecTransf} applies to $\Ll_R$ as well, hence we can conclude at once that ${\rm Spec}(\widehat H) = {\rm Spec}(\widehat{\rho(H)})$.
Furthermore, since $U$, hence also $\Pi_\pm$, are ultra-local, we can be much sharper:
\begin{equation}\label{Eq:BB2}
{\rm Spec}(\widehat H) = {\rm Spec}\big (\Pi_\pm \widehat{\rho(H)}\Pi_\pm \big ).
\end{equation}
The important conclusion is that, if $\widehat H$ displays a gapless boundary spectrum, so does $\widehat{\rho(H)}$ as well as the individual components $\Pi_\pm \widehat{\rho(H)}\Pi_\pm$, where the latter are viewed as acting on the appropriate truncated Hilbert spaces.

The above statement assures us that the bulk-boundary correspondence transfers if the cuts are clean. In realistic conditions, however, we need to add a boundary term to \eqref{Eq:CutHam}, $\widetilde H = \sum_{\bm x,\bm x' \in \Ll_R} \hat b_{\bm x, \bm x'} \otimes |\bm x \rangle \langle \bm x' |$, with $\hat b_{\bm x,\bm x'} =0$ for $\bm x$ and $\bm x'$ far away from the boundary, to take into account relaxations of or damages to the lattice during the cutting process. Let us recall that if $H$ is a topological condensed matter system from the classification table, then the bulk-boundary correspondence principle holds if and only if $\widetilde H$ enjoys the same symmetries as $H$. Given \eqref{Eq:SpecTransf}, we can state at once that:
\begin{equation}\label{Eq:BB3}
{\rm Spec}(\widehat H + \widetilde H) = {\rm Spec}\big (\Pi_\pm \big (\widehat{\rho(H)} + \rho(\widetilde H) \big ) \Pi_\pm \big ).
\end{equation}
Hence, as long as the boundary potential of the classical system is of the form $\Pi_+\rho(\widetilde H) \Pi_+ \oplus \Pi_-\rho(\widetilde H)  \Pi_-$, {\em i.e.} it respects the $U(1)$ symmetry, the bulk-boundary principle transfers. Note that no assumption was made above about the translation invariance of the system, hence the statements apply also in the presence of strong disorder as long as it respects proper symmetries. 

We reached the main conclusion of the section: {\em Let $H$ be a Hamiltonian from the classification table of topological condensed matter systems. Then $\Pi_+\rho(H)\Pi_+$ displays identical bulk-boundary principle as the original Hamiltonian, provided the boundary conditions do not mix the $\Pi_\pm$ sectors and preserve the original symmetries, as transferred through $\rho$.} 

\begin{figure}
\begin{center}
\includegraphics[clip,width=\linewidth]{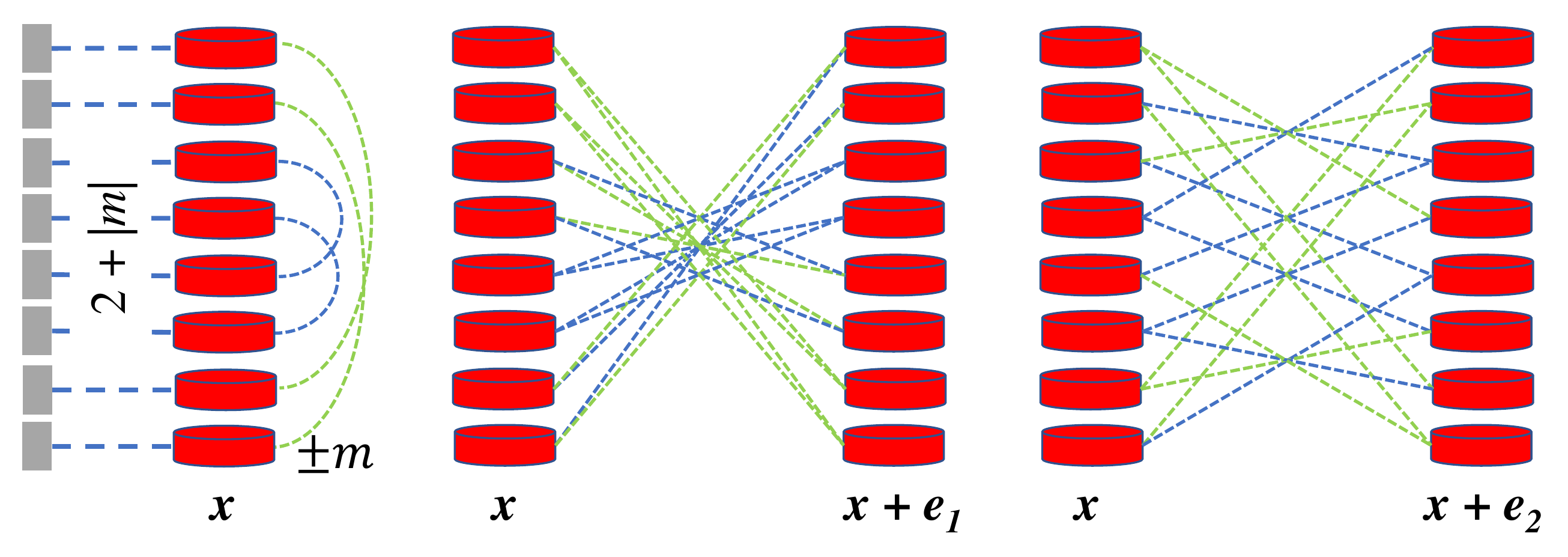}
\caption{\small Schematic representation of the elementary couplings in the realization of the dynamical matrix \eqref{Eq:MappedExample2} corresponding to the generating model of the DIII-AII class. In the figure, $\bm e_j$'s are the primitive vectors of the Bravais lattice in 2D. All couplings have the strength $b=\pm 0.5$, with blue(green) denoting the positive(negative) coupling. The on-site coupling is indicated first panel. The edge spectrum for this model in the topological phase is shown in Fig.~\ref{Fig:2DEdgeSpectra} (c) \& (d).} 
\label{Fig:MappedHam2}
\end{center}
\end{figure}

\section{Filtering the topological sectors}

We describe here one concrete way to excite modes only in one of the $\Pi_\pm$ sectors. The lattice $\Ll$ is considered generic in this section, hence the discussion covers infinite lattices as well as lattices with an edge or with defects. 

The time varying generalized forces can be specified by functions $\bm F_t: \Ll \rightarrow \CM^{2N}$, $\bm F_t \in \ell^2(\Ll,\CM^{2N})$, such that the evaluation at some $\bm x \in \Ll$ gives the instantaneous generalized forces on the degrees of freedom at $\bm x$, $\bm F_t(\bm x) =\big  (F_{\bm x}^1(t), \ldots,F_{\bm x}^{2N}(t)\big )^T$.
When such driving forces are present, the equations of motion in time and frequency domains become:
\begin{equation}
\ddot{\bm Q_t} = -D \bm Q_t + \bm F_t, \quad [ D - \omega^2 ] \bm Q(\omega) =  \bm F(\omega),
\end{equation}
respectively. The solution belongs to one and only one of the $\Pi_\pm$ sectors if and only if $\Pi_\mp \bm F(\omega) = 0$, respectively. We reached the main conclusion of the section: {\em Starting from a generic force field $\bm F_t$, one can excite and drive modes from one and only one of the two $\Pi_\pm$ sectors by simply re-working the force field to:
\begin{equation}
\bm F_t \rightarrow \int_{0}^\infty {\rm Re}\big [\Pi_\pm \bm F(\omega) e^{\imath \omega t}\big ]d \omega,
\end{equation}
respectively.}

As an example, let us consider the case where we start with only one force $F_0 \cos(\omega t)$, acting on the degree of freedom $\alpha \in \{1,\ldots,N\}$ located at vertex $\bm x_0$. It is understood that $\omega$ belongs to the spectrum of the dynamical matrix. In the standard basis, this force field is written as $\bm F_t = F_0 \cos(\omega t) \, \xi_\alpha \otimes |\bm x_0 \rangle$, where $\xi_{\alpha}$ is a unit vector with $1$ in the $\alpha^{th}$ position. To excite only the modes in the $\Pi_+$ sector, we need to modify this force field to $\bm F_t \rightarrow F_0 \, {\rm Re}\big [ e^{\imath \omega t}\Pi_+ \xi_\alpha \big ] \otimes |\bm x_0 \rangle$, which translates into:
\begin{equation} 
\bm F_t \rightarrow F_0 \big [ \cos(\omega t) \xi_\alpha  - \sin(\omega t) \xi_{N+\alpha} \big ] \otimes |\bm x_0 \rangle.
\end{equation}
This tells us that all we have to do is to apply the same generalized force also on the degree of freedom $N+\alpha$ of the same site, and de-phase the force by a quarter of the period. Let us point out that, if $x_0$ happens to be at the edge of the system, this procedure will excite the appropriate edge modes.

\begin{figure}
\begin{center}
\includegraphics[clip,width=\linewidth]{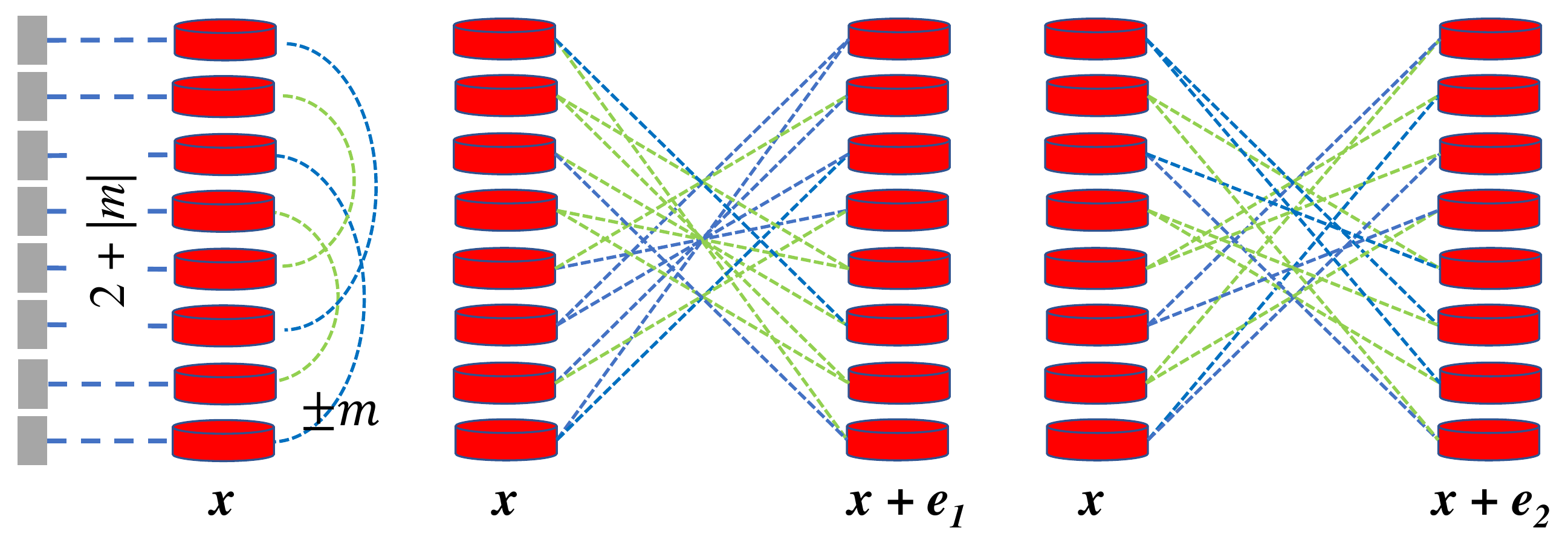}
\caption{\small Schematic representation of the elementary couplings in the realization of the dynamical matrix \eqref{Eq:MappedExample3} corresponding to the generating model of the C class. In the figure, $\bm e_j$'s are the primitive vectors of the Bravais lattice in 2D. All couplings have the strength $b=\pm 0.5$, with blue(green) denoting the positive(negative) coupling. The on-site coupling is indicated first panel. The edge spectrum for this model in the topological phase is shown in Fig.~\ref{Fig:2DEdgeSpectra} (e).} 
\label{Fig:MappedHam3}
\end{center}
\end{figure}

\section{Examples}

The map is applied here to the generating models of the topological classes in 2 and 3 dimensions highlighted in Table~\ref{Tab-ClassTable}. They are all taken from \cite{RSFL2010}. One goal is to verify explicitly all our theoretical predicitions, particularly the identity \eqref{Eq:mappedsymmetry}, for both bulk and half-space Hamiltonians. Another goal is to provide concrete physical representations of the mapped Hamiltonians which turn out to be quite complex, despite the fact that the models are minimal. Let us point out that, while the explicitly worked out examples are periodic, disordered versions of the models can be straightforwardly obtained by considering random fluctuations in the model parameters.  

We will use Pauli's matrices $\sigma_{i}$ together with $\sigma_0 = \IM_2$, as well as the $4 \times 4$ Gamma matrices:
\begin{equation}
\Gamma_{i=1,2,3} ={\scriptsize \begin{pmatrix} 0 & \sigma_i \\ \sigma_i & 0 \end{pmatrix}}, \quad \Gamma_4=\imath {\scriptsize \begin{pmatrix} 0  & -\sigma_0 \\ \sigma_0 & 0 \end{pmatrix}}, \quad \Gamma_0 = {\scriptsize \begin{pmatrix} \sigma_0 & 0 \\ 0 &- \sigma_0 \end{pmatrix}},
\end{equation}
and $\Gamma_{\rm E}=\Gamma_0 \Gamma_2 \Gamma_4$. Since all the Hamiltonians are translational invariant, we adopt the notation from \eqref{Eq:TranslationalH}, involving the shift operators $S_j|\bm x\rangle = |\bm x + \bm e_j \rangle$, where $\bm e_j$'s are the primitive vectors of a Bravais lattice.

\begin{figure}
\begin{center}
\includegraphics[clip,width=\linewidth]{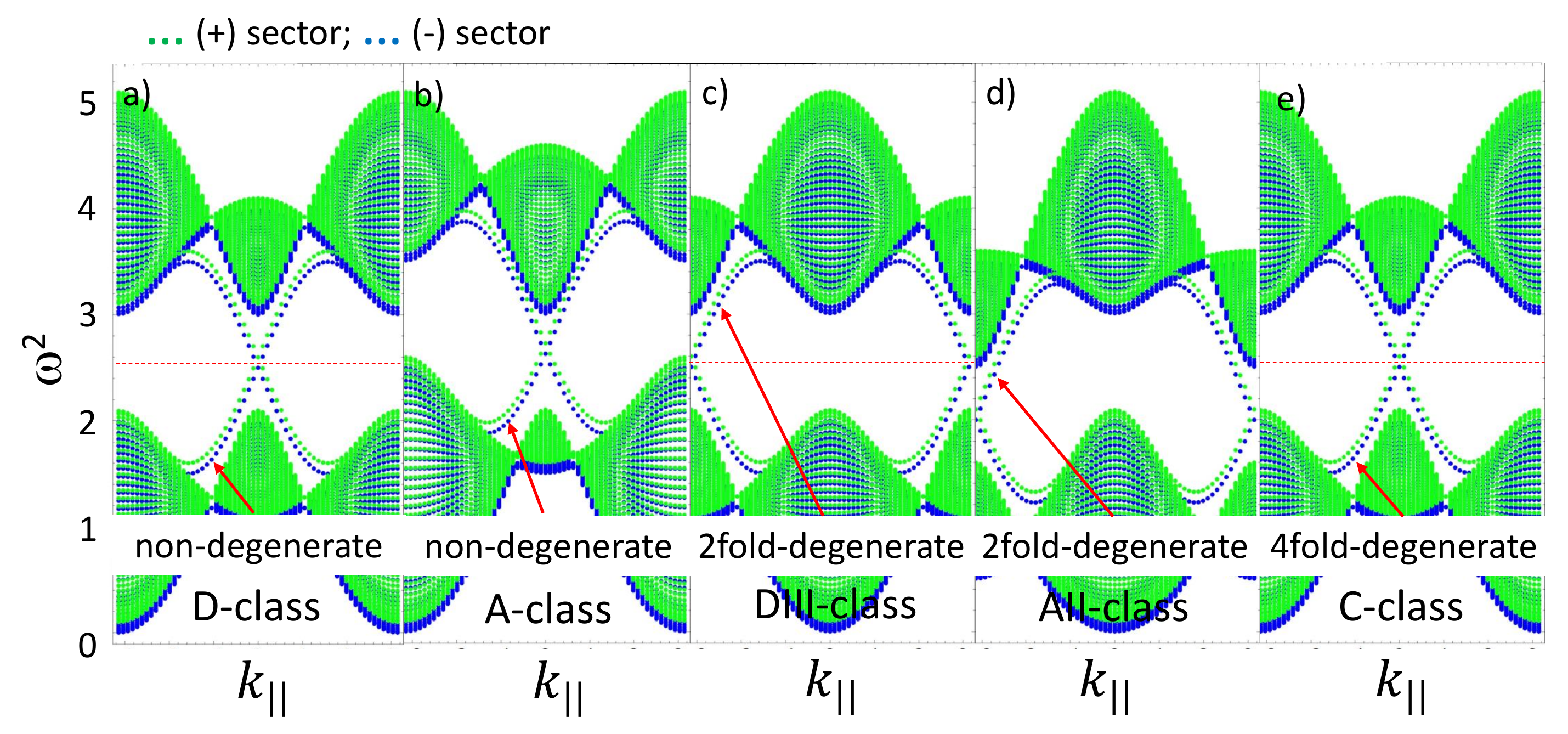}
\caption{\small The spectra of the $d=2$ mapped Hamiltonians with a clean edge, rendered as functions of quasi-momentum parallel to the edge ($\omega$ is the pulsation of the modes). The data was generated with \eqref{Eq:MappedExample1} ($m=-0.5$) for D-class, \eqref{Eq:MappedExample1} ($m=-0.5$) plus a PH-symmetry breaking potential for A-class, \eqref{Eq:MappedExample2} ($m=0.5$) for DIII-class, \eqref{Eq:MappedExample2} ($m=0.5$) plus a PH-symmetry breaking potential for AII-class and \eqref{Eq:MappedExample3} ($m=-0.5$) for C-class. The presence of mirror symmetry relative to the midgaps is indicated by the red-dotted lines and the degeneracy of the edge modes is specified in each panel. The spectra corresponding to the $\Pi_\pm$-sectors have been artificially shifted.} 
\label{Fig:2DEdgeSpectra}
\end{center}
\end{figure}

\noindent{\bf Dimension 2.} {\em D-A Classes.} The strong topological phases from these classes in $d=2$ are classified by topological invariants that take integer values. Chern insulators and topological fermionic excitations in spinless $p_x+ip_y$ BdG superconductors are condensed matter examples from classes A and D, respectively. At the single particle level, the only difference between class A and D is the presence of a particle-hole symmetry $\Theta_{\rm PH}H \Theta_{\rm PH}^{-1}=-H$ for the latter, with $\Theta_{\rm PH}$ anti-unitary and $\Theta_{\rm PH}^2=+1$. All topological phases from these two classes can be generated from the Hamiltonian:
\begin{equation}\label{Eq:2DADClass}
H = \tfrac{1}{2\imath} \sum_{j=1,2} \sigma_j \otimes (S_j - S_j^\dagger) + \sigma_3 \otimes \big ( m + \tfrac{1}{2}\sum_{j=1,2}(S_j + S_j^\dagger) \big ). 
\end{equation}
The phase diagram of \eqref{Eq:2DADClass} consists of a topological phases with Chern number $-1$, for $m\in (-2,0)$, and $+1$, for $m\in (0,2)$, as well as of a trivial phase for $m \notin [-2,2]$. Throughout, the Chern numbers always refer to the lower bands. By staking \eqref{Eq:2DADClass}, one can obtain any topological phase from class A or D. The PH-symmetry is implemented by $\Theta_{\rm PH} = (\sigma_1 \otimes \IM) \Kk$, as one can verify that $\Theta_{\rm PH} \, H \, \Theta_{\rm PH}^{-1} = -H$. 

The mapped Hamiltonian is given by: 
\begin{align}\label{Eq:MappedExample1}
\rho(H) = &  -\tfrac{1}{2}\sum_{j=1,2} \rho(\imath \sigma_j) \otimes (S_j - S_j^\dagger) \\ \nonumber
 & \quad + \rho(\sigma_3) \otimes \big ( m + \tfrac{1}{2}\sum_{j=1,2}(S_j + S_j^\dagger) \big ),
\end{align}
which can be used to attain the dynamical matrix $D=\rho(H) + (2+|m|) \mathbb{I}_4$. The transferred PH-symmetry operator is $\tilde{\Theta}_{\rm PH} = \rho(\sigma_1 \otimes \IM) \, J \Kk$ and we verified numerically that \eqref{Eq:mappedsymmetry} holds and $\tilde \Theta_{\rm PH}^2=+1$. 

\begin{figure}
\begin{center}
\includegraphics[clip,width=0.8\linewidth]{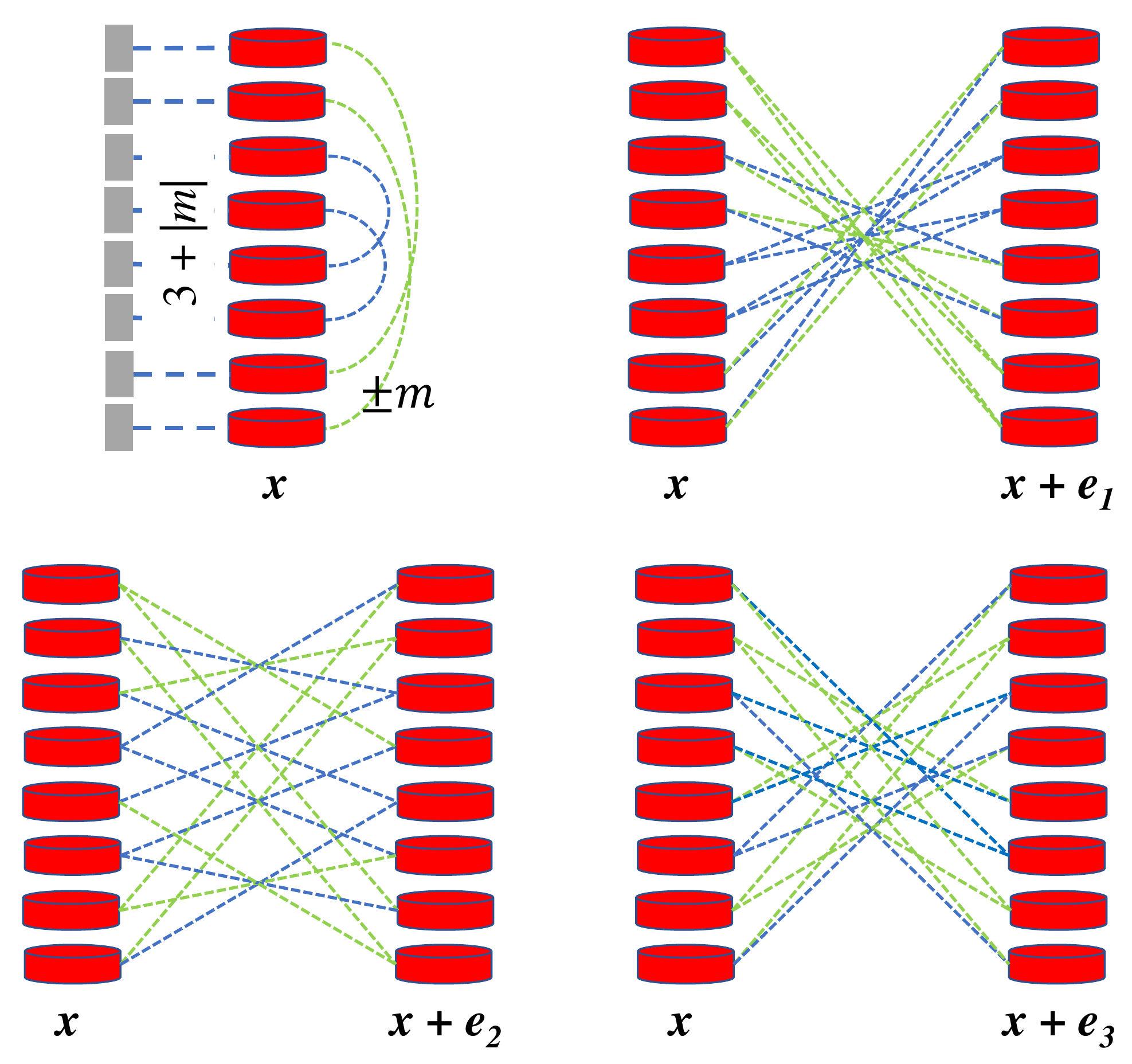}
\caption{\small Schematic representation of the elementary couplings in the realization of the dynamical matrix \eqref{Eq:MappedExample4} corresponding to the generating model of the DIII-AIII-AII classes. In the figure, $\bm e_j$'s are the primitive unit vectors of the Bravais lattice in 3D. All couplings have the strength $b=\pm 0.5$, with blue(green) denoting the positive(negative) coupling. The on-site coupling is indicated first panel. The edge spectrum for this model in the topological phase is shown in Fig.~\ref{Fig:3DEdgeSpectra} (a), (b) \& (c).} 
\label{Fig:MappedHam4}
\end{center}
\end{figure}

 \noindent {\em DIII-AII Classes.} These systems correspond to topological fermionic excitations in triplet p-wave superconductors and to quantum spin-Hall insulators, respectively. At the single particle level, the only difference between AII and DIII classes is that the latter displays an additional PH-symmetry. In $d=2$, all these topological phases are classified by $\ZM_2$ and the simplest common representative model is:
\begin{equation}\label{Eq:Example2}
H = \tfrac{1}{2\imath} \sum_{j=1,2} \Gamma_j \otimes (S_j - S_j^\dagger) + \Gamma_4 \otimes \big ( m + \tfrac{1}{2}\sum_{j=1,2}(S_j + S_j^\dagger) \big ).
\end{equation}
The phase diagram of \eqref{Eq:Example2} consists of a topological phase for $m\in (-2,2)$ and of a trivial phase in rest. Upon adding trivial bands, any Hamiltonian from DIII or AII classes can be adiabatically connected to the phases of \eqref{Eq:Example2}. TR-symmetry is implemented by $\Theta_{\rm TR} = (\imath \Gamma_2 \otimes \IM)\Kk$, $ \Theta_{\rm TR}^2 = -1$, and PH-symmetry by 
$\Theta_{\rm PH} = (\imath \Gamma_1 \Gamma_3 \Gamma_4 \otimes \IM)\Kk$, $\Theta_{\rm PH}^2 = 1$. The combined TR and PH symmetries result in an additional CH-symmetry, given by $W_{\rm CH} = \Gamma_0 \otimes \IM$. 

The mapped Hamiltonian is given by: 
\begin{align}\label{Eq:MappedExample2}
\rho(H) = & -\tfrac{1}{2} \sum_{j=1,2} \rho(\imath \Gamma_j) \otimes (S_j - S_j^\dagger) \\ \nonumber
& \quad + \rho(\Gamma_4) \otimes \big ( m + \tfrac{1}{2}\sum_{j=1,2}(S_j + S_j^\dagger) \big ),
\end{align}
which can be used to attain the dynamical matrix $D=\rho(H) + (2+|m|) \IM_8$. The transferred symmetry operators are $\tilde{\Theta}_{\rm TR} = \rho(\imath\Gamma_2 \otimes \IM) \, J \Kk$ and $\tilde{\Theta}_{\rm PH} = \rho(\imath \Gamma_1 \Gamma_3 \Gamma_4 \otimes \IM ) \, J  \Kk$ respectively, and we have verified numerically that they satisfy the identity \eqref{Eq:mappedsymmetry}. Furthermore, the transferred CH-symmetry operator is $ \tilde{W}_{\rm CH} = \rho(\Gamma_0 \otimes \IM)$. 
  
\noindent {\em C Class.} In $d=2$, the topological phases from this class are classified by $2\ZM$ and the generating model is:
\begin{equation}\label{Eq:Example3}
H = \tfrac{1}{2\imath} \big ( \Gamma_1 \otimes (S_1 - S_1^\dagger) + \Gamma_3 \otimes (S_2 - S_2^\dagger) \big )+  \imath \Gamma_{\rm E} \otimes \big ( m + \tfrac{1}{2}\sum_{j=1,2}(S_j + S_j^\dagger) \big ).
\end{equation}
The phase diagram of \eqref{Eq:Example3} consists of the topological phases with Chern number $-2$ for $m\in (-2,0)$ and $+2$ for $m \in (0,2)$, as well as of a trivial phase if $m \notin (-2,2)$. By staking \eqref{Eq:Example3}, one can generate all the topological phases from class C. The PH-symmetry operators is $\Theta_{\rm PH} = (\Gamma_2 \Gamma_4 \otimes \IM)\Kk $, with  $\Theta_{\rm PH}^2 = -I$. 

The mapped Hamiltonian is:
\begin{align}\label{Eq:MappedExample3}
\rho(H) = & -\tfrac{1}{2} \rho(\imath \Gamma_1) \otimes (S_1 - S_1^\dagger) - \tfrac{1}{2} \rho(\imath \Gamma_3) \otimes (S_2 - S_2^\dagger) \\ \nonumber 
& \quad +  \rho(\imath \Gamma_{\rm E}) \otimes \big ( m + \tfrac{1}{2}\sum_{j=1,2}(S_j + S_j^\dagger) \big ),
\end{align}
which can be used to attain the dynamical matrix $D= \rho(H) + (2+|m|) \mathbb{I}_8$. The transferred PH-symmetry operation is given by $\tilde{\Theta}_{\rm PH} = \rho(\Gamma_2 \Gamma_4 \otimes \IM) \, J  \Kk$ and we verified numerically that it satisfies the identity \eqref{Eq:mappedsymmetry}. 

\noindent{\em Physical Models.} The symmetry operators as well as the mappings in \eqref{Eq:MappedExample1}, \eqref{Eq:MappedExample2} and \eqref{Eq:MappedExample3} are all computed explicitly in the supplemental material. They can be used in conjunction with the procedure explained in section~\ref{Sec:MechRez} to translate \eqref{Eq:MappedExample1}, \eqref{Eq:MappedExample2} and \eqref{Eq:MappedExample3} into physical models, as shown in Figs.~\ref{Fig:MappedHam1}, \ref{Fig:MappedHam2} and \ref{Fig:MappedHam3}, respectively. Since the models are periodic, we only need to specify $\hat{d}_{\bm x,\bm x}$ and $\hat{d}_{\bm x,\bm x+\bm e_j}$, $j=1,2$.

\noindent{\em Edge Modes.} The rendering of the spectra of $\widehat{\rho(H)}$ as function of quasi-momentum parallel to the boundary is shown in Fig~\ref{Fig:2DEdgeSpectra}, for all the above models. The overlapping spectra corresponding to the $\Pi_{\pm}$ sectors have been artificially displaced for visualization. The numerical calculations confirm that $\Pi_+ \widehat{\rho(H)}\Pi_+$ displays topological edge modes with the expected symmetries and degeneracies.

The edge spectra in Fig~\ref{Fig:2DEdgeSpectra} (a) \& (b) correspond to class D and A respectively. Examples of the class D and A quantum model correspond to the $p_{x}+ \imath p_{y} $ superconductors supporting Majorana-zero modes and Chern insulators respectively. Our classical implementation of class D realizes a mid-gap edge mode of a symmetric bulk spectrum, whereas class A is the classical analogue of two copies of Chern insulators. The class AII and DIII edge spectra shown in Fig~\ref{Fig:2DEdgeSpectra} (d) \& (c) corresponds to the classical implementation of the $\mathbb{Z}_2$ topological invariant, which in the quantum models corresponds to time reversal symmetric 2D topological insulators and topological superconductors. To our knowledge these classes have yet to realized in classical meta-materials. Fig~\ref{Fig:2DEdgeSpectra} (e) represents the classical analogue of the 2D class C model which has as yet to be observed in a quantum system.

\begin{figure}
\begin{center}
\includegraphics[clip,width=\linewidth]{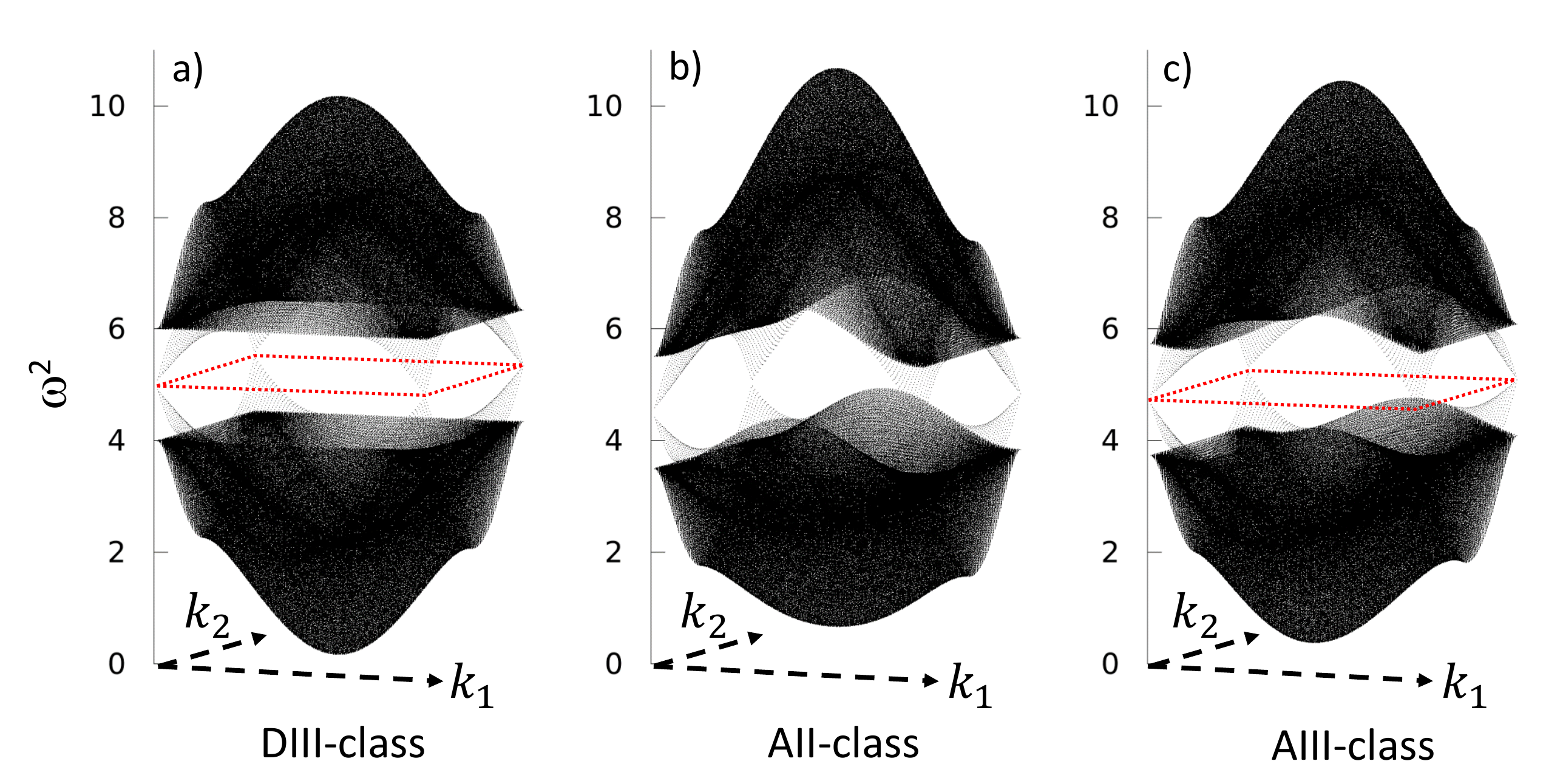}
\caption{\small The spectra of the $d=3$ mapped Hamiltonian with a clean surface, rendered as functions of quasi-momenta parallel to surface. The data was generated with \eqref{Eq:MappedExample4} ($m=2$) for DIII-class, \eqref{Eq:MappedExample4} ($m=2$) plus a PH-symmetry breaking potential for AII-class and \eqref{Eq:MappedExample4} ($m=2$) plus a TR-symmetry breaking potential for AIII-class. All systems display one surface Dirac cone. The presence of mirror symmetry relative to the midgaps is indicated by the red-dotted plane.} 
\label{Fig:3DEdgeSpectra}
\end{center}
\end{figure}

{\bf Dimension 3.} {\em DIII-AIII-AII Classes.} We start with the DIII class, which is the most symmetric. The topological phases from this class in $d=3$ are classified by $\ZM$ and the generating model is:
\begin{equation}\label{Eq:Example4}
H = \tfrac{1}{2\imath} \sum_{j=1}^3 \Gamma_j \otimes (S_j - S_j^\dagger) + \Gamma_4 \otimes \big ( m + \tfrac{1}{2}\sum_{j=1}^3(S_j + S_j^\dagger) \big ).
\end{equation}
The phase diagram of \eqref{Eq:Example4} consist of the topological phases with winding number $+1$ for $m \in (-3,-1)\cup (1,3)$, and with winding $-2$ for $m\in (-1,1)$, and a trivial phase in rest. By staking \eqref{Eq:Example4} we can generate all the topological phases from DIII-class in $d=3$. 

The TR-symmetry is  implemented by $\Theta_{\rm TR} = (\imath \Gamma_2 \otimes \IM)\Kk$ with $\Theta_{\rm TR}^2 = - 1$, the PH-symmetry is implemented by $\Theta_{\rm PH} = (\imath \Gamma_1\Gamma_3\Gamma_4 \otimes \IM)\Kk$, while the CH-symmetry operator can be expressed as $W_{\rm CH} = \Gamma_0 \otimes \IM$.  The AII class can be generated from the DIII class by breaking PH-symmetry but keeping the TR-symmetry. The AII class contains the well known 3D topological insulator which exhibits and odd number of surface Dirac cones and the bulk magneto-electric effect. Breaking the TR-symmetry of the DIII class gives the AIII class, which is marked by the presence of an integer number of surface Dirac cones. 

The mapped Hamiltonian takes the form:
\begin{align}\label{Eq:MappedExample4}
\rho(H) = & -\tfrac{1}{2} \sum_{j=1}^3 \rho(\imath \Gamma_j) \otimes (S_j - S_j^\dagger) \\ \nonumber 
& \quad + \rho(\Gamma_4) \otimes \big ( m + \tfrac{1}{2}\sum_{j=1}^3(S_j + S_j^\dagger) \big ).
\end{align}
which can be used to attain the dynamical matrix $D={\bm \rho}(H) + (3+|m|) \IM_8$. The transferred TR-symmetry and PH-symmetry operators are given by $\tilde{\Theta}_{\rm TR} = \rho(\imath \Gamma_2 \otimes \IM)\, J \Kk$ and $\tilde{\Theta}_{\rm PH} = \rho(\imath \Gamma_1 \Gamma_3 \Gamma_4 \otimes \IM)\, J \Kk$, respectively, while chiral symmetry is implemented by the operator $ \tilde{W}_{\rm CH} = \rho(\Gamma_0 \otimes \IM)$. We have verified numerically that these operators satisfy the identity \eqref{Eq:mappedsymmetry}. The symmetry operations as well as the mappings in \eqref{Eq:MappedExample4} are all computed explicitly in the supplemental material. They can be used in conjunction with the procedure explained in section~\ref{Sec:MechRez} to translate \eqref{Eq:MappedExample4} in a physical model, as shown in Figs.~\ref{Fig:MappedHam4}.

The spectrum of \eqref{Eq:MappedExample4} with a clean surface and appropriate symmetry breaking potentials is reported in Fig.~\ref{Fig:3DEdgeSpectra}. The calculations were done with $\hat D(k)$ computed from \eqref{Eq:TranslationalH}. The results confirm the presence of topological singularities, as well as the predicted symmetries and degeneracies. The surface states shown in Fig~\ref{Fig:3DEdgeSpectra} (a), 
(b) \& (c) obey a Dirac dispersion. Fig~\ref{Fig:3DEdgeSpectra} (b) is the classical realization of the 3D topological insulators which 
belong to the AII class quantum model. Quantum models from the DIII and AIII classes have yet to be realized experimentally, however, our results provide a possible route towards designing classical analogues of these systems in meta-materials.   

Let us conclude by pointing out that all the above periodic models can be transformed in disordered models respecting all the symmetries by replacing the mass parameter $m$ with a diagonal operator $\sum_{\bm x} (m+\delta_{\bm x})|\bm x \rangle \langle \bm x|$, where $\delta m_{\bm x}$ are randomly generated from some interval centered at zero. This amounts to replacing $m$ in all our diagrams by a site dependent value $m_{\bm x}$. The examples do not need to stop at the disorder case, however, since now we have a practical way to implement the topological amorphous condensed matter system studied in \cite{BourneJPA2018} with passive meta-materials. This will be an alternative to the topological amorphous system already demonstrated with active meta-materials \cite{MitchellNP2018}. 

\section{Conclusion}

The algorithm introduced in section~\ref{Sec:Algorithm} enables one to make identical classical copies of any topological condensed matter system from the classification table. This provides an alternative venue to probe, confirm and apply the rich and exotic physical properties of the topological phases. Hence, our algorithm may resolve some outstanding open problems such as resolving the critical regime and measuring the critical exponents of the topological transitions in the presence of disorder, verifying firsthand the robustness of the edge/surface states against disorder, confirming the quantized bulk physical responses, such as the magneto-electric effect in AII-class, and stabilization and manipulations of majorana states.

\acknowledgments{Both authors acknowledge financial support from the W.M. Keck Foundation.}

\end{document}